\documentclass[11pt,showpacs,preprintnumbers,amsmath,amssymb,prd,nofootinbib,superscriptaddress]{revtex4-2}

\usepackage{dcolumn}
\usepackage{bm}
\usepackage{ifpdf}
\usepackage{hyperref}
\usepackage{bm}
\usepackage{xcolor,color,graphicx,graphics,physics}
\usepackage[spanish,english]{babel}
\usepackage[latin1]{inputenc}
\usepackage[OT1]{fontenc}
\usepackage{latexsym,amssymb,amsmath,amsfonts, slashed}
\usepackage{makeidx}
\usepackage{epsfig,subfigure}
\usepackage{natbib}
\usepackage{epstopdf}
\usepackage{mathrsfs}
\usepackage{hyperref}
\hypersetup{colorlinks=true, linkcolor=blue, citecolor=blue, urlcolor=blue}
\usepackage{enumerate}

\usepackage{braket}
\usepackage{fixmath}

\everymath{\displaystyle}
\usepackage{graphicx}

\graphicspath{ {./Imagens/} }

\usepackage[T1]{fontenc}
\usepackage{amsmath}
\usepackage{amssymb}
\usepackage{graphicx}
\usepackage{xcolor}

\newcommand{\bea}{\begin{eqnarray}}
\newcommand{\eea}{\end{eqnarray}}

\newcommand{\orcid}[1]{\href{https://orcid.org/#1}{\includegraphics[width=10pt]{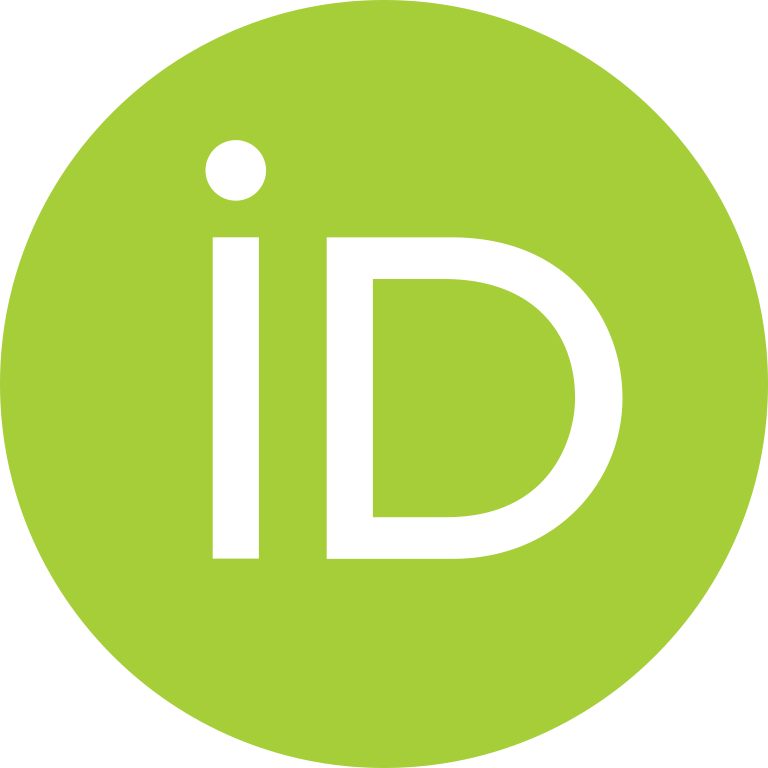}}}

\begin{document}

\title{Barrow holographic dark energy: a path to reconstructing $f(R,T)$ gravity}

\author{P. S. Ens \orcid{0000-0002-7274-2308}}
\email{peter@fisica.ufmt.br  }
\affiliation{Programa de P\'{o}s-Gradua\c{c}\~{a}o em F\'{\i}sica, Instituto de F\'{\i}sica, Universidade Federal de Mato Grosso, Cuiab\'{a}, Brasil}

\author{A. F. Santos \orcid{0000-0002-2505-5273}}
\email{alesandroferreira@fisica.ufmt.br}
\affiliation{Programa de P\'{o}s-Gradua\c{c}\~{a}o em F\'{\i}sica, Instituto de F\'{\i}sica, Universidade Federal de Mato Grosso, Cuiab\'{a}, Brasil}

\begin{abstract}	

In this paper, we investigate one of the established methods for reconstructing modified gravity models from a dark energy model, with the aim of discovering relationships between these theories. In this study, we focus on the $f(R,T)$ modified gravity theory, where $R$ denotes the Ricci scalar and $T$ represents the trace of the energy-momentum tensor. We employ Barrow's holographic dark energy model, derived from fractal surfaces of black holes, to investigate the reconstruction process. The numerical results are subsequently presented for various infrared cutoffs, such as the Hubble horizon, future event horizon, and Granda-Oliveros cutoff.
\end{abstract}	

\maketitle


\section{Introduction}
	
One of the most important discoveries in the last few decades is the late accelerated expansion of the universe, challenging the theory of general relativity and its extensive tests. The attempts made to explain this phenomenon have given birth to a vast set of propositions, many of them seeking ways to modify or add possibly missing parts to general relativity. Some consider the possibility of an energy permeating space, called dark energy, with extraordinary characteristics such as negative pressure and non-conservation. The most well-known candidate for dark energy is the cosmological constant, which induces an exponentially accelerated expansion. Considering an isotropic and homogeneous universe filled with this energy and dark matter, one arrives at the $\Lambda CDM$ cosmological model. Although general relativity and the $\Lambda CDM$ model are adequate to explain most of the observed phenomena \cite{Will2014}, they do not offer a satisfactory explanation for the universe's expansion, since it is known that it does not evolve exponentially and varies over different epochs. Since constant dark energy cannot reproduce such observations, there has been an extensive search for dynamic forms.
	
One possibility for dynamic dark energy is adopted from the holographic principle \cite{Suss, Bou}, which is used to describe black hole thermodynamics \cite{Bekenstein1972, PhysRevD.7.2333, HAWKING1974, PhysRevD.13.191}. This theory aims to study systems based on the characteristics of their boundaries. This concept can be applied to the universe as a whole, derived from the dynamics of a defined infrared cutoff. Here, a form of holographic dark energy is adopted from the idea of event horizons with fractal structures proposed by Barrow \cite{Barrow, PhysRevD.102.123525, NOJIRI2022136844}. New equations of motion emerge from the presence of this energy, which we solved numerically for different infrared (IR) cutoffs: the Hubble radius, the future event horizon, and the Granda-Oliveros cutoff.  Some of the constraints for this model are discussed in \cite{barrow1, saridakis2020bhde}. A similar study has been conducted on Tsallis holographic dark energy in \cite{Zadeh2018}. Barrow dark energy has also been studied in the context of thermodynamics in \cite{Saridakis_2020, Saridakis2021, LUCIANO2023101256}.

In addition to considering an unobserved dark entity, some believe that general relativity needs corrections that would explain the expansion of the universe and are only relevant on cosmological scales. One way to achieve such corrections is by modifying the Lagrangian density that describes general relativity. In general relativity, the Lagrangian is proportional to the Ricci scalar, $R$. The most common and straightforward proposal for modification involves making the Lagrangian a function $f(R)$ \cite{R}, which may include non-linear terms. Taking it a step further, some theories introduce an additional level of interaction between curvature and the universe's components, leading to modifications such as $f(R,T)$, where $T$ represents the trace of the energy-momentum tensor. Reviews on $f(R,T)$ modified gravity can be found in \cite{Harko:2011kv, galaxies2030410, Bhattacharjee2020, SINGH2022168958, Nagpal2021, PhysRevD.95.123536, PhysRevD.90.044031}.

Many models of dark energy and modified gravity offer intriguing results and can solve some observed problems. However, they often create new issues such as unrealistic fine-tuning and instabilities. These positive and negative characteristics can be found across various models, even among different concepts such as dark energy and modified gravity. It is possible to relate both concepts through their equations of state (EoS). A reconstruction of some $f(R,T)$ models from Bekenstein-Hawking holographic dark energy is presented in \cite{HOUNDJO_2012}. An analytical solution for the reconstruction of two $f(R,T)$ models, relating their effective state parameters to those of Tsallis holographic dark energy \cite{Tsallis2013}, considering an evolution of the Hubble parameter, is obtained in \cite{ZUBAIR2022169068}. Other reviews on the reconstruction of dark energy and modified gravity can be found in \cite{doi:10.1142/S0218271806009704, doi:10.1142/S0218271812500034}. In this work, we consider a sufficiently general form for $f(R,T)$, associating it with Barrow's holographic dark energy in a manner that facilitates plotting and estimation of the terms within the $f(R,T)$ model. In this context, we investigate the reconstruction of $f(R,T)$ theory based on the Barrow holographic dark energy, considering three different infrared cutoffs.

This paper is organized as follows. In Section \ref{II}, we introduce the Barrow holographic dark energy, derive the Friedmann equation, and calculate the equation of state parameter and the density parameter for three different infrared cutoffs: the Hubble horizon, the future event horizon, and the Granda-Oliveros cutoff. In Section \ref{III}, we present $f(R, T)$ gravity, derive an equation of state parameter, and develop a reconstruction of $f(R,T)$ theory considering the Barrow holographic dark energy. We explore this reconstruction for the Hubble horizon, the future event horizon, and the Granda-Oliveros cutoff, discussing some numerical results. Finally, in Section \ref{IV}, we provide some concluding remarks.

\section{Holographic dark energy} \label{II}

In this section, the holographic dark energy scenario is briefly presented. Holographic dark energy is an interesting construct that describes dark energy, originating from the holographic principle \cite{Suss, Bou}. In this context, Barrow holographic dark energy is introduced \cite{Barrow}. The state parameter and the density parameter will be investigated for different infrared (IR) cutoffs, including the Hubble horizon, the future event horizon, and the Granda-Oliveros cutoff.

Let us consider an isotropic and homogeneous universe with spacetime geometry defined by the flat Friedmann-Robertson-Walker (FRW) metric, given as
\bea
 ds^2=-dt^2+a(t)^2(dr^2+r^2d\Omega^2),\label{1}
\eea 
where $a(t)$ is the scale factor. This universe is filled with a perfect fluid whose energy-momentum tensor is given by 
\bea
T_{\mu\nu}=(\rho+p)u_\mu u_\nu+pg_{\mu\nu}\label{2}
\eea
with $\rho$, $p$ and $u_\mu$ being the energy density, pressure and the four-velocity of the fluid, respectively. With these ingredients, the Einstein field equations lead to the Friedmann equation
\begin{equation}\label{FriedmannEquation}
3H^2=\kappa(\rho_m+\rho_{DE}),
\end{equation}
where $\rho_m$ and $\rho_{DE}$ denote the energy density of dark matter (DM) and dark energy (DE), respectively. We assume that DM and DE do not interact with each other. Therefore, they satisfy the standard conservation equations, i.e.,
\bea
\dot{\rho}_m+3H\rho_m&=&0,\label{4}\\
\dot{\rho}_{DE}+3H(1+\omega_{DE})\rho_{DE}&=&0\label{5}
\eea
with $\omega_{DE}=p_{DE}/\rho_{DE}$ being the equation of state (EoS) parameter of DE.  

Taking the time derivative of Eq. \eqref{FriedmannEquation} and using Eqs. (\ref{4}) and (\ref{5}), we obtain
\begin{equation}
2\dot{H}=-\kappa(\rho_m+\rho_{DE}+\omega_{DE}\rho_{DE}).\label{6}
\end{equation}
Defining the density parameter as $\Omega_i\equiv\rho_i/\rho_c=\frac{\kappa \rho_i}{3H^2}$, where $\rho_c$ is the critical energy density and $\kappa=8\pi G$, Eq. (\ref{6}) can be written as
\begin{equation}\label{razaohpontoh2}
\frac{\dot{H}}{H^2}=-\frac{3}{2}(1+\omega_{DE}\Omega_{DE}).
\end{equation}
It has been used that $\Omega_m+\Omega_{DE}=1$.

	Using this relation, it is possible to determine the state parameter and the density parameter given one or the other. We can find a solvable equation for the density parameter and differentiate it with respect to time ($t$), e-fold number ($\ln a$), redshift ($z$), or matter density. Then
\begin{equation}\label{derivativesrelation}
\frac{d\Omega(t)}{dt}=H(\ln a)\frac{d\Omega(\ln a)}{d\ln a}=-H(z)(1+z)\frac{d\Omega(z)}{dz} = -3H(\rho_m)\rho_m\frac{d\Omega(\rho_m)}{d\rho_m}.
\end{equation}
In a general form, it can be expressed as
\begin{equation}
\frac{d\Omega}{dt}=\frac{d\ln a}{dt}\frac{d\Omega}{d\ln a}=H\Omega'(\ln a).
\end{equation}
From the definition of the density parameter, it becomes
\begin{equation}\label{omegadifflna}
H\Omega'_{DE}(\ln a)=\partial_t \left(\frac{\kappa \rho_{DE}}{3H^2}\right)=\frac{k}{3}(\dot{\rho}_{DE}H^{-2}-2\rho_{DE}H^{-3}\dot{H})
\end{equation}
with a dot representing the time derivative and a prime denoting the derivative with respect to the function argument. This relation can be solved for a given energy density, here taken from the holographic principle.
	
	The holographic principle postulates that the description of a system  can be based on its surface properties rather than its volume, suggesting that entropy is bounded to be proportional to the area of the boundary. As a result, the energy density, constrained by the entropy as $L^3 \rho_{DE}^{3/4}\leq S^{3/4}$, is also confined to the area, that is,
\begin{equation}
\rho_{DE} \leq \frac{B}{L^2},
\end{equation}
where $B$ is a parameter to be defined, $L$ is the infrared cutoff representing the longest distance of the system. We choose the value of $\rho_{DE}$ that saturates the inequality.
	
	In this study, we consider the Barrow holographic dark energy model \cite{Barrow}. Barrow demonstrated how the boundary area of a black hole can exhibit significantly different values when viewed as a fractal. In this scenario, the system's entropy becomes $S=S_{BH}^{1+\Delta/2}$, with $\Delta$ representing the intricacy of the surface, where $\Delta=0$ indicates perfect smoothness and $\Delta=1$ represents maximum complexity. From this entropy expression, we derive Barrow holographic dark energy given as
\begin{equation}
\rho_{DE}=\rho_{BHDE}=BL^{\Delta-2}.
\end{equation}
	It is possible to insert this energy density into Eq. \eqref{omegadifflna}, but finding its solution and the state parameter will necessitate defining an infrared cutoff for the holographic dark energy. Now, three different IR cutoffs will be considered.
	

\subsection{Hubble Horizon}

\begin{figure}[htb]
\begin{center}
	\includegraphics[height=5cm]{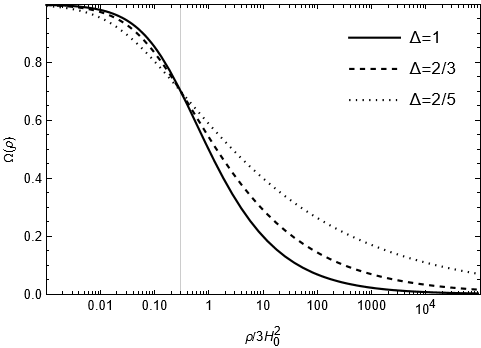} \quad
	\includegraphics[height=5cm]{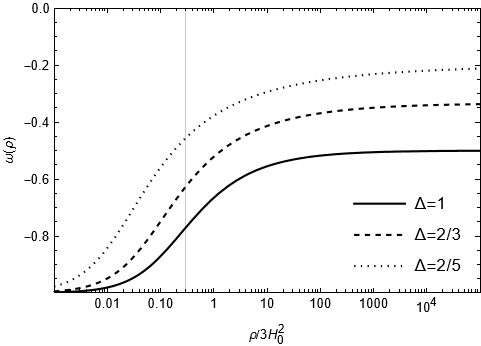}
\caption{Plot of numeric solutions for the density parameter (left) and EoS parameter (right), using $\Omega_{DE0}=0.7$, with  different values of the $\Delta$ parameter. The vertical lines in the plots represent $\rho= \rho_{m0}$, i.e. the present time.} 
\label{fig1}
\end{center}
\end{figure}

In this subsection, the Hubble horizon cutoff is introduced \cite{Hsu, Xu, Sri}, followed by an analysis of the cosmological behavior of the universe. The Hubble horizon cutoff is defined as $L=H^{-1}$. Then, the energy density of DE becomes 
\bea
\rho_{DE}=BH^{2-\Delta}.
\eea
Taking its time derivative and using the energy conservation relation Eq. (\ref{5}), we obtain
\begin{equation}
\omega_{DE}=-\frac{(2-\Delta)\dot{H}}{3H^2}-1.
\end{equation}
From Eq. \eqref{razaohpontoh2}, the state parameter can be expressed as a function of the density parameter, i.e.,
\begin{equation}
\omega_{DE}=\frac{\Delta}{(2-\Delta)\Omega_{DE}-2}.\label{15}
\end{equation}
Considering Eqs. \eqref{omegadifflna} and \eqref{razaohpontoh2}, the follwing relation for the density parameter is obtained
\begin{equation}
\Omega'_{DE}=3\Delta\Omega_{DE}\left(\frac{\Omega_{DE}-1}{(2-\Delta)\Omega_{DE}-2}\right).
\end{equation}
From here, we can numerically find the solution. We can also change the dependency of the density parameter to $\rho_m$, resulting in the following differential equation:
\begin{equation}
\Omega'_{DE}=-\frac{\Delta\Omega_{DE}}{\rho_m}\left(\frac{\Omega_{DE}-1}{(2-\Delta)\Omega_{DE}-2}\right).
\end{equation}
This will later allow the calculation of a solution for $f(R,T)$ models in scenarios involving dust matter. In Figure \ref{fig1}, the numerical solutions for the density parameter and state parameter as a function of energy density are shown, considering the Hubble horizon as the IR cutoff. In this context, our results indicate that an accelerating expansion of the universe is permitted when the Hubble cutoff is considered. Furthermore, for the Hubble cutoff, acceleration can only be achieved with non-zero delta $\Delta$ values. In all these scenarios, exponential expansion occurs in the distant future. Delta values below 2/3 lead to a transition from deceleration to acceleration at some point in the past.

\subsection{Future Event Horizon}\label{futurehorizon}

	Since the Hubble radius is not a suitable event horizon given that we can observe objects beyond it, many prefer to explore holographic theories using other distances, such as the future event horizon \cite{Sad}. In this case, the infrared cutoff is defined as 
\begin{equation}
L=R_h=a(t)\int_{t}^{\infty}\frac{dt}{a(t)}
\end{equation}
with $\dot{R}_h=HR_h-1$.	

	Using the future event horizon as the infrared cutoff, the Barrow holographic dark energy density becomes $\rho_{DE}=BR_h^{\Delta-2}$, and its time derivative  is given by
\begin{equation}\label{futurehorizondensityderivative}
\dot{\rho}_{DE}=\rho_{DE}(\Delta-2)H(1-F),
\end{equation}
where 
\begin{equation}
F=\left(\frac{3\Omega_{DE}H^{\Delta}}{\kappa B}\right)^{\frac{1}{2-\Delta}}
\end{equation}
with $\Omega_{DE}=\frac{\kappa}{3H^2}\rho_{DE}$. From the energy conservation equation, i.e. $\dot{\rho}_{DE}+3H(1+\omega_{DE})\rho_{DE}=0$,  and using Eq. \eqref{futurehorizondensityderivative}, the EoS parameter is written as
\begin{equation}
\omega_{DE}=-1-\frac{\Delta-2}{3}(1-F).\label{21}
\end{equation}
	
	Substituting these ingredients into Eq. \eqref{omegadifflna} and using Eq. \eqref{razaohpontoh2}, the differential equation for the density parameter is given as
\begin{equation}\label{densityparameterdif}
\Omega'_{DE}=\Omega_{DE}(1-\Omega_{DE})[1-F(\Delta-2)+\Delta].
\end{equation}
	
To accurately calculate the above equation, it is necessary to address the Hubble parameter within $F$. This can be achieved by using the Friedmann equation and expressing it as follows:
\begin{equation}
H^2=\frac{\kappa}{3}(\rho_{m0}(1+z)^3+\rho_{DE})=H_0^2\Omega_{m0}(1+z)^3+H^2\Omega_{DE}=H_0^2\frac{1-\Omega_{DE0}}{1-\Omega_{DE}}(1+z)^3,
\end{equation}
where the suffixes $0$, such as $m0$ and $DE0$ refer to values at the current epoch. 

\begin{figure}[htb]
\begin{center}
	\includegraphics[height=5cm]{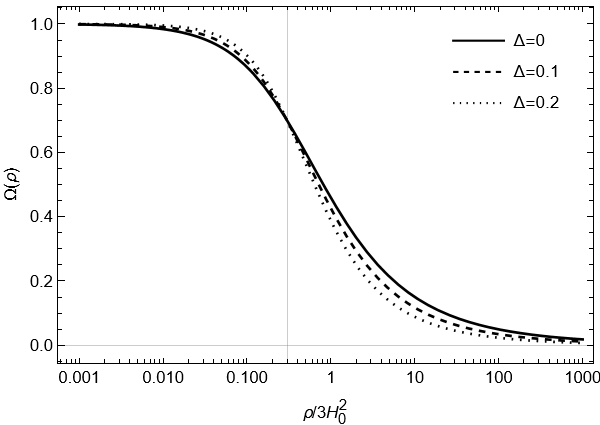} \quad
	\includegraphics[height=5cm]{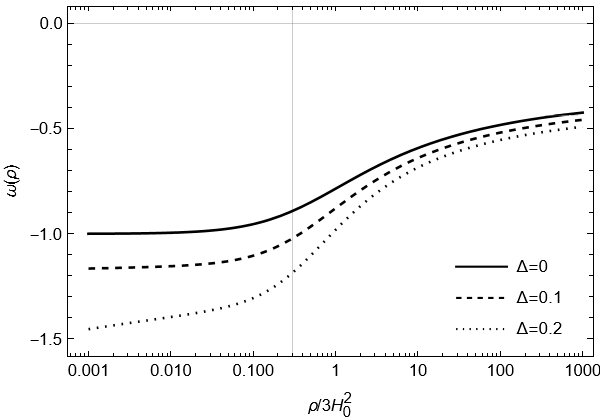}
\caption{Plot of numeric solutions for the density parameter (left) and EoS parameter(right), using $\Omega_{DE0}=0.7$ and $H_0=67$, with different values of $\Delta$.} 
\label{fig2}
\end{center}
\end{figure}

The numeric solutions for $\Omega_{DE}$ and $\omega_{DE}$, considering the future event horizon as the infrared cutoff, are shown in Figure \ref{fig2}. For the future event horizon cutoff, acceleration always occurs regardless of the value of delta. In contrast to the Hubble case, this parameter significantly influences the behavior in the distant future.

It is important to note that employing the method of first solving $H(z)$, differentiating the Friedmann equation, and then calculating the density and EoS parameter yields the same results as presented above.
	
\subsection{Granda-Oliveros cutoff}

	The future event horizon represents a potential causality violation, since its calculation requires knowledge of the scale factor value in the future \cite{LI20041}. L. N. Granda and A. Oliveros proposed a purely mathematical approach as a cutoff to avoid this issue \cite{GO1, GO2}. The Barrow holographic dark energy with the Granda-Oliveros (GO) cutoff has been investigated in \cite{Oliveros2022}. With this cutoff, $L=\left(\alpha H^2+\beta\dot{H}\right)^{-1/2}$, we can readily determine the energy density of dark energy as follows
	\begin{equation}
	\rho_{DE}=B\left(\alpha H^2+\beta\dot{H}\right)^{\frac{2-\Delta}{2}}
	\end{equation}
	which leads to
	\begin{equation}
	\frac{\dot{H}}{H^2}=\frac{1}{\beta}\left((\Omega_{DE}H^\Delta)^\frac{2}{2-\Delta}-\alpha\right)\equiv F.
	\end{equation}
	Here, $B=3m_p^2$ has been chosen, and this expression is denoted as $F$ for simplicity. Inserting this result into Eq. \eqref{omegadifflna} and substituting $\dot{\rho}_{DE}=3m_p^2(2H\dot{H}+3H^3\Omega_m)$ from Friedmann and energy conservation equations, we obtain
	\begin{equation}
	\Omega_{DE}'=(1-\Omega_{DE})(3+2F).
	\end{equation}
	
	Using Eq. \eqref{razaohpontoh2} the EoS parameter is given as
	\begin{equation}
	\omega_{DE}= -\frac{2F+3}{3\Omega_{DE}}.\label{27}
	\end{equation}
	
\begin{figure}[htb]
	\begin{center}
	\includegraphics[height=5cm]{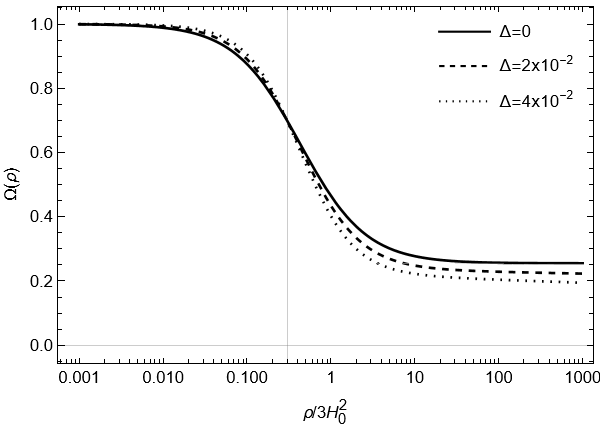} \quad
	\includegraphics[height=5cm]{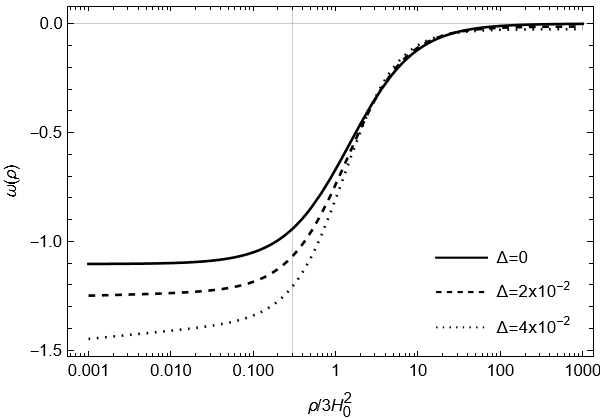}
	\caption{Plot of numeric solutions for the density parameter (left) and EoS parameter (right), using $\Omega_{DE0}=0.7$, $H_0=67$, $\alpha=0.93$ and $\beta=0.45$, with different values of $\Delta$.} 
	\label{fig3}
	\end{center}
	\end{figure}

	Similar to the previous cases, the density parameter and EoS parameter are numerically determined, as shown in Figure \ref{fig3}.	 The GO cutoff preserves the characteristics of the future event horizon case for both the current and distant future epochs. However, in all scenarios, it induces the transition from deceleration to acceleration more prominently than the Hubble case.
	
\section{Reconstructing an $f(R,T)$ model from Barrow holographic dark energy}\label{III}

In this section, the main objective is to reconstruct the $f(R, T)$ gravity within the Barrow holographic dark energy model. First, a brief introduction to the standard $f(R, T)$ gravity is provided. Then, it is proposed to reconstruct the theory considering different event horizons.

This gravitational theory consists of a generalization of the Einstein-Hilbert action of general relativity. In this case, the Ricci scalar $R$ in the general relativity action is replaced by a function $f$ that depends on both the Ricci scalar $R$ and the trace of the energy-momentum tensor $T$. The action that describes $f(R,T)$ gravity is given by
\begin{equation}
S=\int \left[f(R,T) + \mathcal{L}_m\right] \sqrt{-g} d^4x,\label{action}
\end{equation}
where $g$ represents the determinant of the metric and $\mathcal{L}_m$ is the Lagrangian describing the matter content. A review on $f(R,T)$ gravity can be found in \cite{PhysRevD.84.024020, Myrzakulov2012-zy}. Several studies have been developed considering this theory of gravity. For example, reconstructions of cosmological models within $f(R,T)$ theory have been investigated in \cite{HOUNDJO_2012, Jamil_2012, Sharif2014-md}, while perturbations, constraints, and the Palatini formulation are presented in \cite{PhysRevD.87.103526, Alvarenga_2013, Wu_2018}.

Varying the action (\ref{action}) with respect to the metric and considering a perfect fluid as the matter content, with a Lagrangian defined by $\mathcal{L}_m=-p$, the field equations are given as
\begin{equation}
f_R(R,T)R_{\mu\nu}-\frac{1}{2}f(R,T)g_{\mu\nu}+(g_{\mu\nu}\Box-\nabla_\mu\nabla_\nu)f_R(R,T)=T_{\mu\nu}+f_T(R,T)T_{\mu\nu}+pf_T(R,T)g_{\mu\nu},
\end{equation}
where $f_R(R,T)=\frac{\partial f(R,T)}{\partial R}$, $f_T(R,T)=\frac{\partial f(R,T)}{\partial T}$ and $T_{\mu\nu}$  is the energy-momentum tensor which is defined as
\begin{equation}\label{eq:2}
     T_{\mu\nu}=-\frac{2}{\sqrt{-g}}\frac{\delta(\sqrt{-g}\mathcal{L}_{m})}{\delta g^{\mu\nu}}.
\end{equation}

Here, let's choose one of the simplest models of $f(R, T)$ gravity, which consists of the usual general relativity term, the Ricci scalar $R$,  plus an $f(T)$ correction, i.e., $f(R,T)=R+2f(T)$. Then, the field equations become
\begin{equation}
R_{\mu\nu}-\frac{1}{2}Rg_{\mu\nu}=T_{\mu\nu}+2f_T(T)T_{\mu\nu}+2pf_T(T)g_{\mu\nu}+f(T)g_{\mu\nu}.
\end{equation}

Taking an isotropic and homogeneous universe described by the FRW metric Eq. (\ref{1}) filled with a perfect fluid whose energy-momentum tensor is given in Eq. (\ref{2}), leads to the modified Friedmann equations
\bea\label{friedmannfRT1}
3H^2=\rho+2(\rho+p)f_T(T)+f(T) &\equiv& \rho + \rho_{f(T)}\\
-2\dot{H}-3H^2=p-f(T) &\equiv& p + p_{f(T)}.
\eea
Here, $\rho_{f(T)}$ and $p_{f(T)}$ represent the modified energy density and pressure associated with the $f(T)$ contribution. These quantities are defined as follows
\bea
\rho_{f(T)}&=&2(\rho+p)f_T(T)+f(T) ,\\
p_{f(T)}&=&-f(T).
\eea

It is important to note that from these elements, the equation of state parameter induced by the function $f(T)$ is defined as
\begin{equation}
\omega_{f(T)}\equiv \frac{p_{f(T)}}{\rho_{f(T)}}=-\frac{f(T)}{2(\rho+p)f_T+f(T)}\label{36}
\end{equation}
 and the deceleration parameter is given by
\begin{equation*}
	q\equiv-1-\frac{\dot{H}}{H^2}=\frac{\rho+3p+(\rho+p)2f_T-2f(T)}{2\rho+4(\rho+p)f_T+2f(T)}.
	\end{equation*}
It is important to highlight that the interpretation of the deceleration parameter aligns with that of the state parameter.

With these results, the main objective now is to reconstruct the $f(R,T)$ theory from Barrow holographic dark energy. For this, let's assume that the equation of state (EoS) parameter given in Eq. (\ref{36}) is equal to the EoS parameter of dark energy $( \omega_{DE})$ in a context where the matter content is dust. The trace of the energy-momentum tensor for this matter content is $T=\rho$. Then, by taking $\omega_{f(T)} = \omega_{DE}$, we obtain
\begin{equation}\label{eqdiffrt}
f_\rho=-\frac{f(\rho)}{2\rho}\left(1+\frac{1}{\omega_{DE}}\right).
\end{equation}
 It is worth noting that in the specific case of $\omega_{DE}=-1$, this derivative vanishes, leading to a constant $f(\rho)$. Substituting this into Eq. \eqref{friedmannfRT1} recovers the well-known de Sitter evolution.
	
To correctly calculate the solution of $f(\rho)$, we need an initial condition. From the modified Friedmann equations \eqref{friedmannfRT1}, we find
\begin{equation}
f_\rho(\rho_0)=\frac{3H_0^2\Omega_{DE0}}{2\rho_0}-\frac{f(\rho_0)}{2\rho_0}.
\end{equation}
Using this result in Eq. (\ref{eqdiffrt}) at the current time leads to
\begin{equation}
f(\rho_0)=-3H_0^2\Omega_{DE0}\omega_{DE0}.
\end{equation}

In order to continue the reconstruction of $f(R,T)$ theory using Eq. (\ref{eqdiffrt}), different IR cutoffs will be considered, namely the Hubble horizon, the future event horizon, and the Granda-Oliveros cutoff.
	
\subsection{Hubble horizon}

As a first case, the Hubble cutoff is considered. For this cutoff, the EoS parameter is given in Eq. (\ref{15}). Using this result, Eq. (\ref{eqdiffrt}) becomes
\begin{equation}
f_\rho=-\frac{f(\rho)}{2\rho}\left(1+\frac{(2-\Delta)\Omega_{DE}-2}{\Delta}\right),
\end{equation}
with initial condition given by
\begin{equation}
f(\rho_{m0})=-\frac{3H_0^2\Omega_{DE0}\Delta}{(2-\Delta)\Omega_{DE0}-2}.
\end{equation}

\begin{figure}[htb]
\begin{center}
	\includegraphics[height=5cm]{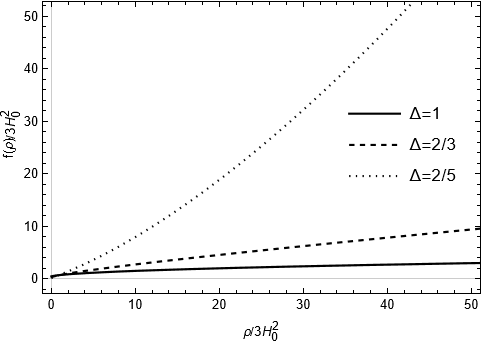} \quad
	\includegraphics[height=5cm]{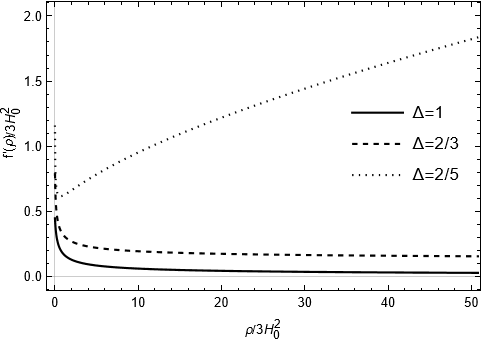}
\caption{Plot of numeric solutions for $f(\rho_m)$ (left) and its derivative (right), using $\Omega_{DE0}=0.7$.} 
\label{fig4}
\end{center}
\end{figure}

\begin{figure}[htb]
	\begin{center}
		\includegraphics[height=5cm]{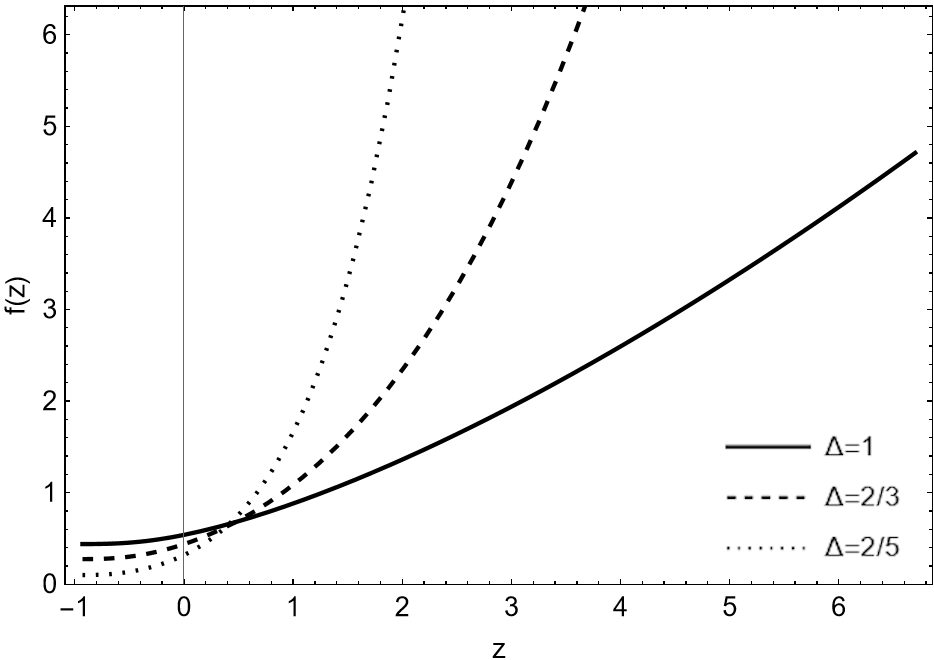}
		\caption{Plot of numeric solution for $f(z)$, using $\Omega_{DE0}=0.7$.} 
		\label{fig77}
	\end{center}
\end{figure}

We can immediately see that for $\Delta=0$, it is not possible to compute $f(\rho)$. However, for $\Delta=2$, we obtain the cosmological constant scenario with $f(\rho)=\rho_{DE0}$. For other values of $\Delta$, using the results of $\Omega$, we calculate the solution for $f(\rho_m)$. This solution for different values of $\Delta$  is shown in Figures \ref{fig4} and \ref{fig77}.

At first glance, the early component of the function ($f_E(\rho)$) appears to follow a power law, i.e.,
\begin{equation}
f_E(\rho)=\alpha\rho^{g(\Delta)},
\end{equation}
where $\alpha$ is a constant and $g(\Delta)$ a function of $\Delta$. For $\Delta$ values close to zero, the slope of the function increases  dramatically and begins to resemble an exponential.
	
Values of $\Delta$ were chosen based on their effect on the function in the past. In this case, $f(\rho)$ appears to be a square root for $\Delta=1$, linear for $\Delta=2/3$, and quadratic for $\Delta=2/5$. This can also be verified by solving the equations for $\Omega_{DE} \ll 1$, providing a hint of the form the function should take. In the far future ($\Omega_{DE} \approx 1$), the function becomes constant. But in all situations, there is a change in behavior from the near past to the future, with a significant positive change in the slope of $f$ in the future and its derivative always being positive. This indicates the presence of other term(s) in the $f(\rho)$ function that dominate at lower $\rho_m$ values. As of now, our best guess for the late component ($f_L(\rho)$) is something like
\begin{equation}
f_L(\rho)=a\ln(\rho^{h(\Delta)}+c) \quad \text{or} \quad f_L(\rho)=b\,\rho^{h(\Delta)} ,
\end{equation}
where $a, b$ and $c$ are constants and $0<h(\Delta)<1$.

In the next subsection, we investigate $f(\rho)$ considering a different horizon cutoff, thus reconstructing the $f(R,T)$ theory from a different perspective.

\subsection{Future Event Horizon}

Here, the future event horizon cutoff is considered. Using the EoS parameter given in Eq. (\ref{21}), we can express Eq. (\ref{eqdiffrt}), which determines the function $f_\rho$, as	
\begin{equation}
f_\rho=-\frac{f(\rho)}{2\rho}\left(1-\frac{3}{3+(\Delta-2)(1-F)}\right).
\end{equation}
The initial condition is given by
\begin{equation}
f(\rho_{m0})=-3H_0^2\Omega_{DE0}\left(-1-\frac{\Delta-2}{3}(1-F_0)\right).
\end{equation}
	
\begin{figure}[htb]
\begin{center}
	\includegraphics[height=5cm]{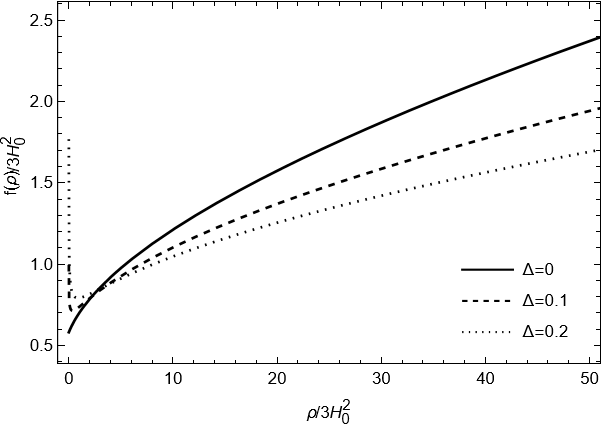} \quad
	\includegraphics[height=5cm]{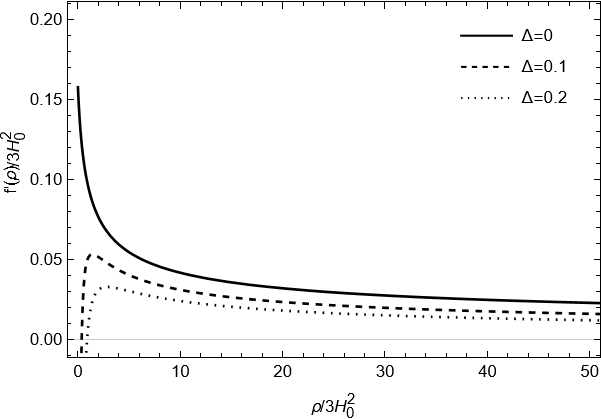}
\caption{Plot of numeric solutions for $f(\rho_m)$ (left) and its derivative (right), using $\Omega_{DE0}=0.7$, $H_0=67$.} 
\label{fig5}
\end{center}
\end{figure}

\begin{figure}[htb]
\begin{center}
\includegraphics[height=5cm]{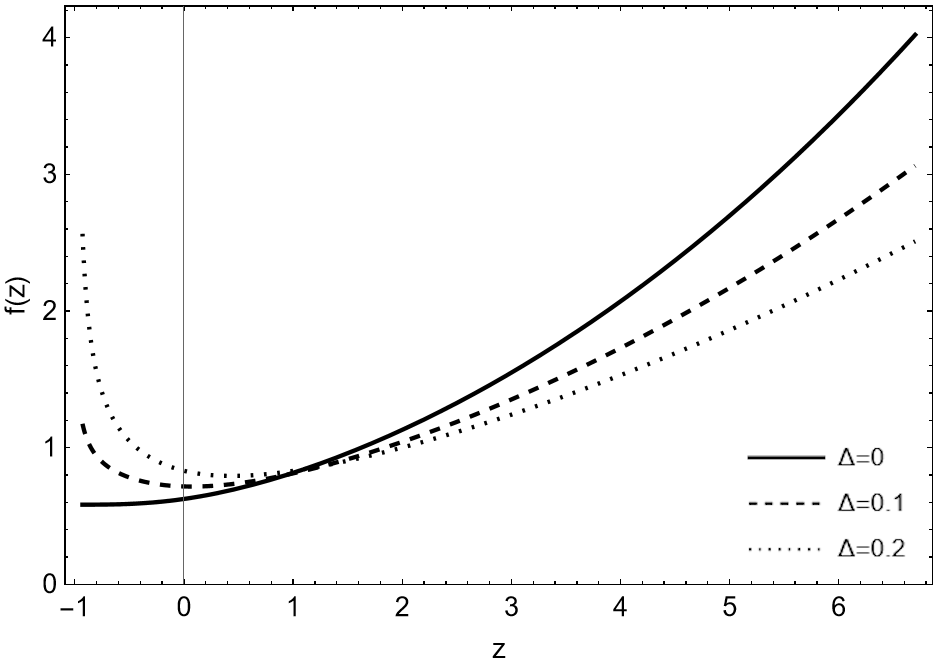}
\caption{Plot of numeric solution for $f(z)$, using $\Omega_{DE0}=0.7$, $H_0=67$.} 
\label{fig55}
\end{center}
\end{figure}

Using the results of $\Omega$, we calculate the solution for $f(\rho_m)$. These results are shown in Figures \ref{fig5} and \ref{fig55}, considering different values of $\Delta$.		

There is clearly a functional transition to $\Delta \neq 0$. In the past, it behaves as a power law, $f_E(\rho) \propto \rho^{g(\Delta)}$, with $g(\Delta) < 1$ inversely proportional to $\Delta$. Toward the present and future, it transitions to $f_L(\rho) \propto \rho^{-h(\Delta)}$, with $h(\Delta)$ also inversely proportional to $\Delta$ and disappearing into $\Delta = 0$. The first derivative $f'(\rho)$ appears to be a function like
\begin{equation}
f'(\rho)\propto\rho^{-g'(\Delta)}-\rho^{-h'(\Delta)}
\end{equation}
where $h'(\Delta) > g'(\Delta)$, and $h'(\Delta)$ vanishes for $\Delta = 0$. This result is largely similar to the one obtained with the Hubble cutoff during the early stages of expansion but deviates significantly as we approach the current era.
	
\subsection{Granda-Oliveros cutoff}

\begin{figure}[htb]
\begin{center}
	\includegraphics[height=5cm]{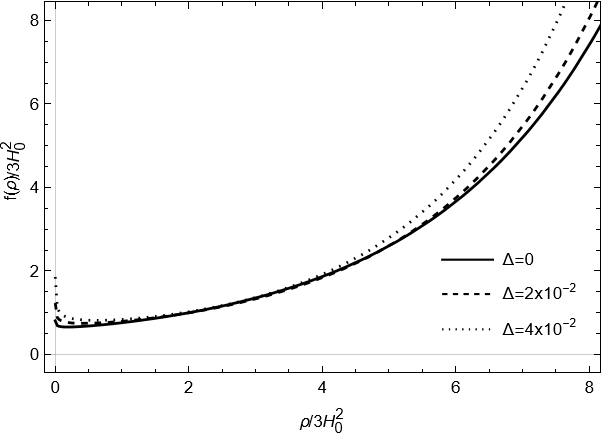} \quad
	\includegraphics[height=5cm]{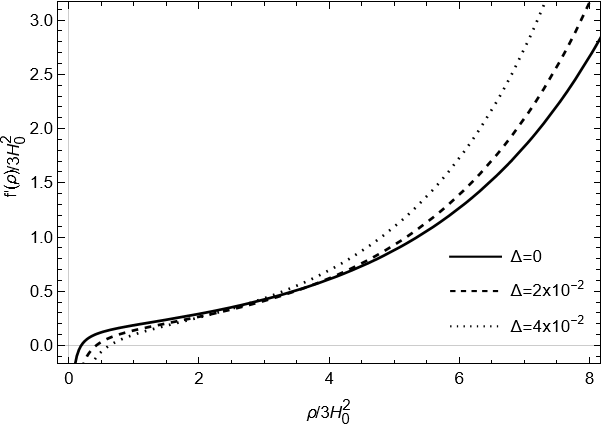}
\caption{Plot of numeric solutions for $f(\rho_m)$ (left) and its derivative (right), using $\Omega_{DE0}=0.7$, $H_0=67$, $\alpha=0.93$ and $\beta=0.45$.} 
\label{fig6}
\end{center}
\end{figure}

\begin{figure}[htb]
	\begin{center}
		\includegraphics[height=5cm]{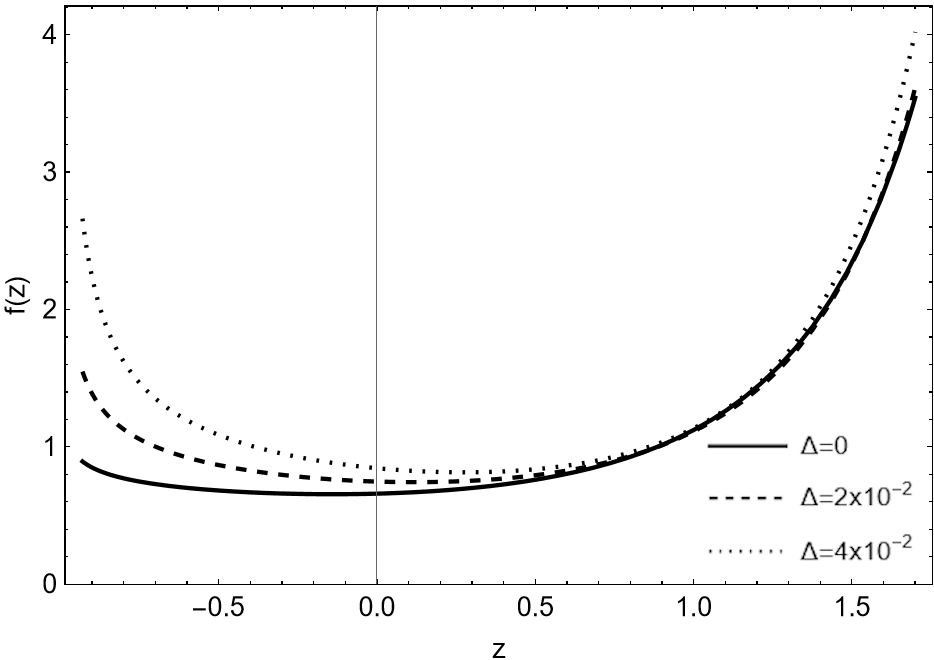}
		\caption{Plot of numeric solution for $f(z)$, using $\Omega_{DE0}=0.7$, $H_0=67$, $\alpha=0.93$ and $\beta=0.45$.} 
		\label{fig7}
	\end{center}
\end{figure}

The third IR cutoff being considered is the Granda-Oliveros cutoff. For this cutoff, the EoS parameter is defined in (\ref{27}). Then, for this case, Eq. (\ref{eqdiffrt}) is expressed as
\begin{equation}
f_\rho=-\frac{f(\rho)}{2\rho}\left(1-\frac{3\Omega_{DE}}{3+2F}\right)
\end{equation}
and its initial condition is
\begin{equation}
f(\rho_{m0})=-3H_0^2\Omega_{DE0}\left(-\frac{3+2F_0}{3\Omega_{DE0}}\right)
\end{equation}
with $F_0=\frac{1}{\beta}\left((\Omega_{DE0}H_0^\Delta)^\frac{2}{2-\Delta}-\alpha\right)$. Using the results of $\Omega$, the solution for $f(\rho_m)$ is found, as exhibited in Figures \ref{fig6} and \ref{fig7}.

From the function and its derivatives, we can observe that it is exponential in the past for any values of $\Delta$, given by
\begin{equation}
f(\rho)\propto e^{g(\Delta)\rho} \quad \text{or} \quad f(\rho)\propto e^{\rho^{g(\Delta)}},
\end{equation}
where $g(\Delta) \propto \Delta$. There is a transition in the present and future that appears to become an inverse power law, $f(\rho) \propto \rho^{-g(\Delta)}$, with $g(\Delta) \propto 1/\Delta$, similar to the one proposed for the future event horizon.  These approximated solutions have been studied in the context of $f(R,T)$ and $f(R,L_m)$ gravity \cite{JEAKEL2024101401, Harko2010}.

For both the future event horizon and Granda-Oliveros cutoffs, we obtain $f(\rho)$ solutions where its influence in later eras is significantly larger than 
$\rho_m$, implying that the Hubble parameter, as derived from the modified Friedmann equations (\ref{friedmannfRT1}), will resemble these solutions. This behavior aligns with the expected trends observed in the state parameters, as discussed in the previous section, showing a transition from power-law expansion to exponential growth and beyond.

\section{Conclusion}\label{IV}

The late-time accelerated expansion of the universe remains one of the most challenging problems of the last few decades. To address this problem, various attempts have been made. These include the proposal of alternative gravity theories and the consideration of exotic forms of matter or energy. This work investigates a dark energy model based on the holographic principle, known as Barrow holographic dark energy.  In this context, we calculate the equation of state parameter and the density parameter of dark energy under three different cutoff conditions: the Hubble horizon, the future event horizon, and the Granda-Oliveros cutoff, with the latter exhibiting the expected behavior while avoiding some of the theoretical issues associated with the other two. Our findings reveal distinct scenarios that describe the accelerated expansion of the universe in the present epoch. Additionally, the Barrow holographic dark energy model is utilized to reconstruct the $f(R, T)$ theory. This reconstruction is analyzed across various infrared cutoffs. The results demonstrate that it is possible to determine the function $f(\rho)$ from the equation of state parameter associated with certain infrared cutoffs. The analysis presented here focuses on a specific $f(R, T)$ model, namely $f(R,T)=R+2f(T)$, under the assumption that the universe's matter content is dust. We obtain numerical results and utilize them to estimate the behavior of the function $f(\rho)$ across various periods of the universe's evolution. Moreover, our results, which demonstrate the reconstruction of a $f(R,T)$ theory, align with observational data indicating a late acceleration of the universe's expansion.

\section*{Acknowledgments}

This work by A. F. S. is partially supported by National Council for Scientific and Technological Develo\-pment - CNPq project No. 312406/2023-1. P. S. E. thanks CAPES for financial support.

\section*{Data Availability Statement}

No data are available because of the nature of the research. This publication is theoretical work that does not require supporting research data.

\global\long\def\link#1#2{\href{http://eudml.org/#1}{#2}}
 \global\long\def\doi#1#2{\href{http://dx.doi.org/#1}{#2}}
 \global\long\def\arXiv#1#2{\href{http://arxiv.org/abs/#1}{arXiv:#1 [#2]}}
 \global\long\def\arXivOld#1{\href{http://arxiv.org/abs/#1}{arXiv:#1}}

\end{document}